\documentstyle[amssymb,prl,aps,epsf]{revtex}
\twocolumn
\tighten

\begin{document}
\draft

\twocolumn[
\hsize\textwidth\columnwidth\hsize\csname@twocolumnfalse\endcsname
\title{ Spontaneous Vortex Phase in the Bosonic RVB Theory}
\author{ Z. Y. Weng$^a$ and V. N. Muthukumar$^b$}
\address{$^a$ Center for Advanced Study, Tsinghua University,
Beijing 100084, China\\
$^b$ Department of Physics, Princeton University,
Princeton, NJ 08544\\
}
\maketitle 
\begin{abstract}
In the description of spin-charge separation 
based on the phase string theory of the $t-J$ model, spinon 
excitations are vortices in the superconducting state. 
Thermally excited spinons destroy phase coherence, 
leading to a new phase characterized by the presence of
{\em free} spinon vortices at temperatures, $T_c<T<T_v$. 
The temperature scale $T_v$ at which holon condensation occurs
marks the onset of {\em pairing amplitude}, and is related to the
spin pseudogap temperature $T^*$. 
The phase below $T_v$, called the spontaneous vortex phase, shows 
novel transport properties before phase coherence sets in at $T_c$. 
We discuss the Nernst effect as an intrinsic 
characterization of such a phase, in 
comparison with recent experimental measurements. 
\end{abstract} 
]

\narrowtext

\section{Introduction}

A unique feature of superconductivity in the high-$T_c$ cuprates is that
phase coherence in electron pairing may occur at a lower temperature
than the temperature at which the pairing amplitude develops
\cite{V-K}. Doped Mott insulators\cite{anderson}
provide a natural explanation: owing to the separation of spin and charge
degrees of freedom, the spin resonating valence bond (RVB) pairing can be
achieved at a much higher temperature; while the RVB pairs, accompanied by
charge backflow, can move around like Cooper pairs at finite doping, 
superconducting condensation is absent until phase coherence between
the pairs is established at a relatively lower $T_c$ \cite{baskaran}.

Recently, we proposed a 
Ginzburg-Landau description of the RVB superconductor \cite{muthu} based
on the phase string theory\cite{zy} of the $t-J$ model. 
In this description, the superconducting order parameter is given by
\cite{remark2} 
\begin{equation}
\Delta ({\bf r})\sim \Delta ^0({\bf r})e^{i\Phi ^s({\bf r})}  \label{dsc}
\end{equation}
where the Cooper-pair{\em \ amplitude} $\Delta ^0\equiv \Delta ^s[\psi
_h^{*}]^2,$ with $\Delta ^s$ being the {\em bosonic} RVB order parameter of
spins and $\psi _h$ being the charge (holon) Bose condensed field. Here
$\Delta ^0\neq 0$ does not directly result in 
superconductivity; rather, it is
the phase factor in (\ref{dsc}) that determines the {\em phase
coherence} of the pairing order parameter, and thereby, $T_c$.

The quantity
$\Phi ^s({\bf r})$ characterizes phase vortices centered around spinons,
since $\Phi ^s({\bf r})\rightarrow \Phi ^s({\bf r})\pm 2\pi $, 
if ${\bf r}$
winds around a spinon excitation continuously in space\cite{muthu}.
Thus, each spinon excitation induces a phase vortex (called spinon
vortex) in the order parameter. Superconducting phase coherence is
destroyed above $T_c$ by the presence
of {\em free} spinon vortices. Below $T_c$, phase coherence is
realized as spinon vortices and
antivortices are bound \cite{muthu}, such that 
$\Phi ^s( {\bf r})$ becomes trivial in (\ref{dsc}). 
In the superconducting phase, single spinon vortices can only be
present at magnetic vortex cores which ensures flux quantization at 
$hc/2e$. 

It should be noted that
the bosonic RVB order parameter $\Delta ^s$ in (\ref{dsc}) is not
related to the
energy gap, in contrast to the usual fermionic RVB
order parameter\cite{baskaran}. It describes (neutral) spin pairing
as characterized by short range
(nearest-neighbor) antiferromagnetic correlations,
$\langle {\bf S }_i\cdot {\bf S}_j\rangle _{NN}=-1/2|\Delta ^s|^2$.
At small doping, $\Delta ^s\neq 0$
covers a temperature regime extended over $1,000$K \cite{zy}.
On such an RVB background, the
Cooper pair amplitude $\Delta _0=\Delta ^s[\psi _h^{*}]^2$ is realized
when the holons (Bose) condense, {\em i.e.},
$ \psi _h\neq 0$. But the temperature $T_v$ at which
this occurs does not necessarily coincide with 
$T_c$ in general. Thus, one may find a temperature regime 
in the {\em normal state}, $T_v>T>T_c$ 
(obviously $T_v$ cannot be lower
than $T_c$ since it is the holon Bose condensation that engenders
phase coherence \cite{muthu}). 

In this paper, we will focus on a normal state with nonzero
(preformed) pairing amplitude. It is distinguished from a conventional 
normal state of strong superconducting fluctuations by the 
presence of free spinon vortices as elementary excitations. 
We call such a phase, a {\em spontaneous vortex phase}. 
We present the
effective Hamiltonian governing the dynamics of spinon vortices. Some
novel transport properties in this state are discussed, and in
particular, we show that free spinon vortices contribute to 
a nontrivial Nernst signal, consistent with the recent measurements of
the Princeton group \cite{ong,ong2,ong1}. 
We argue that a finite $\Delta _0$ controls the pseudogap phenomena,
and present a phase diagram showing the connection between
the superconducting phase and the phase with preformed pairs
(spontaneous vortex phase). 

\section{Spinon Vortices: Elementary Excitations}

\subsection{Spinons as vortices}

The phase $\Phi ^s({\bf r})$ in (\ref{dsc}) is given by\cite{muthu} 
\begin{equation}
\Phi ^s({\bf r})=\int d^2{\bf r}^{\prime }~{\mbox {Im}}\ln \left[
z-z^{\prime }\right] \left[ n_{\uparrow }^b({\bf r}^{\prime })-n_{\downarrow
}^b({\bf r}^{\prime })\right]   \label{Phis}
\end{equation}
where $n_\sigma ^b$ denotes the density operator of bosonic spin-1/2
excitations (spinons) and $z=x+iy$. As noted in the Introduction, 
$\Phi ^s( {\bf r})$ changes by $\pm 2\pi $ when ${\bf r}$ 
winds around a  spinon once. In the ground state, the
spinons are all RVB paired such that $\Phi ^s( {\bf r})$ 
is effectively canceled out. Excited spinons created
by breaking the RVB pairs lead to vortices in the pairing
order parameter (\ref{dsc}) through $\Phi ^s({\bf r})$.

The spinon vortex can be understood from a different perspective. 
The effective Hamiltonian for the holons is given by\cite{zy,muthu} 
\begin{equation}
{H}_h=\frac 1{{2m_h}}\int d^2{\bf r}~h^{\dagger }({\bf r})\left( -i\nabla -%
{\bf A}^s-{\bf A}^e\right) ^2h({\bf r})  \label{hh}
\end{equation}
where $h({\bf r})$ is the bosonic holon field, and 
${\bf A}^e$ is the vector potential
of the external electromagnetic field. 
(Here we set $\hbar=e=c=1$. )
The quantity ${\bf A}^s$ defined by 
\begin{equation}
{\bf A}^s({\bf r})=\frac 12\int d^2{\bf r}^{\prime }~\frac{\hat{{\bf z}}
\times ({\bf r}-{\bf r}^{\prime })}{|{\bf r}-{\bf r}^{\prime }|^2}\left[
n_{\uparrow }^b({\bf r}^{\prime })-n_{\downarrow }^b({\bf r}^{\prime
})\right]~,   \label{as1}
\end{equation}
is related to $\Phi ^s$ by ${\bf A}^s=\frac 12\nabla \Phi ^s$. Note
that
\begin{equation}
\oint_cd{\bf r}\cdot {\bf A}^s=\frac 12\oint_cd{\bf r}\cdot {\bf \nabla \Phi 
}^s\\ \nonumber
=\pm \pi \int_{S\in c} d^2{\bf r}^{\prime }\left[ n_{\uparrow }^b({\bf r}^{\prime })-n_{\downarrow }^b({\bf %
r}^{\prime })\right]
\end{equation}
i.e., ${\bf A}^s$describes a fictitious flux quantized at $\pm \pi $, bound
to each spinon{\bf \ }as seen by holons.

The vortices in $\Phi ^s({\bf r})$ become meaningful when $\Delta ^0\neq 0$
in (\ref{dsc}), {\em i.e.}, when the holons are Bose condensed. 
In terms of the superfluid
density $\rho _h$ and the phase $\phi _h$, the holon condensate 
$\psi
_h({\bf r})\equiv \left\langle h({\bf r})\right\rangle =\sqrt{\rho _h}
~e^{i\phi _h({\bf r})}$, and equation (\ref{hh}) reduces to 
\begin{equation}
{H}_h\approx \frac 1{{2m_h}}\int d^2{\bf r}~\rho _h\left( \nabla \phi _h-%
{\bf A}^s-{\bf A}^e\right) ^2  \label{hh1}
\end{equation}
The corresponding superfluid current operator\cite{muthu} 
is given by
\begin{equation}
{\bf J}=\frac{\rho _h}{m_h}\left[ \nabla \phi _h-{\bf A}^s-{\bf A}^e\right]
~~.  \label{jc}
\end{equation}

Consider a  spinon excited from the RVB background within a loop $c$.
In terms of (\ref{jc}), there must be an induced current vortex surrounding
the excited spinon (setting $\nabla \phi _h={\bf A}^e=0)$ 
\begin{equation}
\oint_c{\bf J}({\bf r})\cdot d{\bf r}=-\frac{\rho _h}{m_h}\oint_c{\bf A}^s(%
{\bf r})\cdot d{\bf r=}\mp \frac{\rho _h}{m_h}\pi .  \label{vs}
\end{equation}
Therefore, each spinon excitation will always be accompanied by a current
vortex in the holon condensate. This current vortex is consistent with
the aforementioned $\pm 2\pi $ vorticities of $\Phi ^s$ in $\Delta $ 
(in the latter {\em a pair }of holon fields are present).

It should be noted that such a vortex cannot be compensated or screened
by $\nabla \phi _h$ in (\ref{jc}) since the latter must satisfy the
single-valued condition $\oint d{\bf r}\cdot \nabla \phi _h=0$, $\pm 2\pi $,
..., instead of $\pm \pi $, the flux quanta carried by spinons in 
${\bf A}^s$.
However, $\nabla \phi _h$ provides a gauge freedom such 
that the {\em sign} of
the vorticity for each minimal current vortex surrounding an excited
spinon in (\ref{vs}) is not related to the spin polarization itself 
\cite{muthu}. For example, consider a spinon at site
$i$, with 
$\oint d{\bf r}\cdot {\bf A}^s=\frac 12\oint d{\bf r}\cdot {\bf 
\nabla \Phi }^s=\pi$. 
Under a singular gauge transformation,
$\psi _h( {\bf r})\rightarrow \psi _h({\bf r})e^{i\theta ({\bf r})}$ 
with 
$\oint d{\bf r}\cdot \nabla \theta =2\pi$ 
centered at $i$, the theory is invariant
if ${\bf A}^s\rightarrow {\bf A}^s-\nabla \theta $ 
($\Phi ^s\rightarrow \Phi ^s-2\theta $), 
with 
$\oint d{\bf r}\cdot {\bf A}^s=\frac 1
2\oint_cd{\bf r}\cdot {\bf \nabla }\Phi ^s\rightarrow -\pi $. 
In the appendix, we give a general proof that spin rotational symmetry is
indeed preserved here.

Finally, we note that
there also exist the usual Kosterlitz-Thouless (KT) like $2\pi $
vortices in $\psi _h.$ But since we will only be interested in the 
temperature regime, 
$T<T_{v\text{ }}$ where holons are condensed, we ignore their effects.
Consequently, in the rest of the paper, $\nabla \phi _h$ will be
set to zero in (\ref{hh1}) and (\ref{jc}), 
with its singular part associated
with spinons being incorporated into ${\bf A}^s$. So, in the latter [(\ref
{as1})] the spin index should be generally construed as the vorticity index
and independent of the actual spin index, as discussed above.

\subsection{Effective Hamiltonian of spinon vortices}

The effective Hamiltonian governing the spinon vortices can be written in
two parts:

\begin{equation}
H_{spinon-vortex}=H_s+H_v  \label{hsv}
\end{equation}
Here, $H_s$ is the mean field Hamiltonian for bosonic spinon 
excitations\cite
{zy},
\begin{equation}
H_s=\sum_{m\sigma }E_m\gamma _{m\sigma }^{\dagger }\gamma _{m\sigma }+const.
\end{equation}
where $\gamma _{m\sigma }$ is related to the bare bosonic operator by the
Bogoliubov transformation 
\begin{equation}
b_{i\sigma }=\sum_mw_{m\sigma }\left( {\bf r}_i\right) \left[ u_m\gamma
_{m\sigma }-v_m\gamma _{m-\sigma }^{\dagger }\right]   \label{bogo}
\end{equation}
with $u_m=1/\sqrt{2}(\lambda _m/E_m+1)^{1/2}$\ and $v_m=1/\sqrt{2}(\lambda
_m/E_m-1)^{1/2}sgn(\xi _m)$; $E_m=\sqrt{\lambda _m^2-\xi _m^2}${\bf \ }
and $\lambda _m=\lambda -\frac{J_h}{J_s}\left| \xi _m\right| $ ($J_h\sim
\delta t$, $J_s\sim J)$\cite{zy}. The single particle wave function 
$w_{m\sigma }\left( {\bf r}_i\right)$ and $\xi _m$ are determined by the
tight binding equation
\begin{equation}
\xi _mw_{m\sigma }\left( {\bf r}_i\right) =-J_s\sum_{j=NN(i)}e^{-i\sigma
A_{ji}^h}w_{m\sigma }({\bf r}_j)  \label{bdg}
\end{equation}
in which $w_{m\sigma }\left( {\bf r}_i\right) =w_{m-\sigma }^{*}\left( {\bf r%
}_i\right) .$\ Note that $\sum\nolimits_cA_{ij}^h=\pi \sum_{l\in c}n_l^h$
describes $\pi $ flux tubes bound to holons ($n_l^h$ denotes the holon
number operator)\cite{zy}. In the holon Bose condensed phase, it is a good
approximation to treat $A_{ij}^h$ as a vector potential for a {\em uniform}
flux perpendicular to the two-dimensional plane, with a field strength 
\begin{equation}
B^h=\frac{\pi \delta }{a^2}
\end{equation}
where $\delta $ is the doping concentration and $a$ is the lattice constant.

Therefore, the tight binding model in (\ref{bdg}) has a
Hofstadter-Landau level structure. In the weak-field (continuum) limit, 
$\xi _m$ is a discrete, dispersionless Landau-level energy, 
with the wave function characterized by a cyclotron length scale
\begin{equation}
a_c=\frac 1{\sqrt{B^h}}=\frac a{\sqrt{\pi \delta }}  \label{cycl}
\end{equation}
Correspondingly the spinon spectrum $E_m$ is discrete as well, and if we restrict
ourselves to the lowest Landau-level (LLL) at the low temperature, the
dispersionless spectrum 
\begin{equation}
(E_m)_{\text{LLL}}\equiv E_s
\end{equation}
with a degeneracy equal to $2\times B^ha^2/2\pi =$ $\delta $ (the prefactor $%
2$ comes from the fact that $\xi _m$ and -$\xi _m$ solutions are degenerate
in $E_m)$ and $E_s\sim \delta J$ \cite{zy} ($\lambda $ in $E_m$ is determined
by the average constraint condition $\left\langle \sum\nolimits_\sigma
b_{i\sigma }^{\dagger }b_{i\sigma }\right\rangle =1-\delta $). In this theory%
\cite{zy}, 
\begin{equation}
E_g=2E_s
\end{equation}
corresponds to the energy scale of the sharp resonance-like peak
observed in the neutron-scattering measurements\cite{bourges}, with its weight
proportional to the degeneracy of the energy levels, $\delta $.

The second term in (\ref{hsv}), $H_v$, describes the interaction 
among spinons due to the
fact that they carry current vortices. 
This term arises from the effective
Hamiltonian for the holons, $H_h$. Setting $\nabla \phi _h={\bf A}^e=0$ 
in (\ref{hh1}), we get
\begin{eqnarray}
{H}_v &=&\frac{\rho _h}{{2m_h}}\int d^2{\bf r}~\left( {\bf A}^s\right) ^2 
\nonumber \\
&=&\int \int d^2{\bf r}_1d^2{\bf r}_2\sum\nolimits_\alpha \alpha n_\alpha ^b(%
{\bf r}_1)V({\bf r}_{12})\sum\nolimits_\beta \beta n_\beta ^b({\bf r}_2)
\label{hv}
\end{eqnarray}
in which 
\begin{eqnarray}
V({\bf r}_{12}) &=&\frac{\rho _h}{{2m_h}}\int d^2{\bf r}\frac{\hat{{\bf z}}%
\times ({\bf r}-{\bf r}_1)}{|{\bf r}-{\bf r}_1|^2}\cdot \frac{\hat{{\bf z}}%
\times ({\bf r}-{\bf r}_2)}{|{\bf r}-{\bf r}_2|^2}  \nonumber \\
&=&-\frac{\pi \rho _h}{{4m_h}}\ln \frac{|{\bf r}_1-{\bf r}_2|}{r_c}
\end{eqnarray}
with $r_c\sim a$.
Using the Bogoliubov transformation (\ref{bogo}) and defining $n_{m\sigma
}^\gamma =\gamma _{m\sigma }^{\dagger }\gamma _{m\sigma }${\bf \ , }one finds
\begin{equation}
\sum\nolimits_\sigma \sigma n_\sigma ^b({\bf r})=\sum_m\left| w_{m\sigma }(%
{\bf r})\right| ^2\sum\nolimits_\sigma \sigma n_{m\sigma }^\gamma
+\sum_{m\neq n}...
\end{equation}
and thus the interaction term can be further rewritten as $H_v=H_v^0+H_v^1,$
with
\begin{figure}[ht!]
\epsfxsize=7.0 cm
\epsfysize=2.5 cm
\centerline{\epsffile{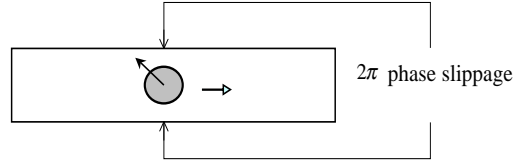}}
\vspace{2mm}
\caption{A $2\pi $ ``phase slip'' takes place in the phase difference
between two edges of the strip when a spinon-vortex passes through the
sample. }
\label{fig:1}
\end{figure}
\begin{equation}
H_v^0=\sum_{\alpha \beta }\alpha \beta \sum_{mn}n_{m\alpha }^\gamma
U_{mn}n_{n\beta }^\gamma   \label{hv0}
\end{equation}
where
\begin{equation}
U_{mn}=\int \int d^2{\bf r}_1d^2{\bf r}_2\left| w_{m\alpha }({\bf r}%
_1)\right| ^2\left| w_{n\beta }({\bf r}_2)\right| ^2V({\bf r}_{12})
\end{equation}
and the non-diagonal part $H_v^1$ describes the scattering among different
states induced by vortex interactions.

The diagonal part $H_v^0$ provides a confining force for spinons in the
superconducting phase. Unlike the conventional KT vortices, there is
a finite core for each spinon-vortex composite in $H_v^0$, within which the
spinon does a cyclotron motion with a core radius $\sim a_c$ determined
by $\left| w_{m\alpha }({\bf r})\right| ^2$. From a  detailed
renormalization group analysis\cite{ming}, it is found that 
$T_c$, the temperature at which
spinon vortices unbind, scales with the spin resonance-like energy $E_g$,
consistent with a simpler physical argument based on the ``core touching''
picture\cite{muthu}, where it has been estimated that the transition occurs
when the number of {\em excited} spinons exceeds the holon number.

\section{Spontaneous vortex phase}

\subsection{Definition}

The spinon-vortex composite described above is a unique elementary
excitation, predicted by the present bosonic RVB theory, in the regime $\psi
_h$ or $\Delta _0\neq 0$ below the holon condensation temperature $T_v.$ As
pointed out in the Introduction, this regime generally includes two phases.
One is the superconducting phase at a lower temperature $T_c$, where phase coherence is
realized when spinon vortices are paired up (``confined''). The second phase
at $T_c<T<T_v$ is a normal state in which unbinding (``deconfined'') free
spinon vortices are present. We define such a state of matter as the
spontaneous vortex phase. One may also regard this phase as a normal state
with a {\em finite} Cooper-pair amplitude $\Delta _0$.

\subsection{Phase rigidity and the Nernst effect}
The existence of free vortices implies that some kind of phase rigidity
persists above $T_c$ in the spontaneous vortex phase. In the following,
we explore some consequences.

Consider a strip sample showing in Fig. 1, in which a
spinon vortex traverses from one end of the sample to the other along the
strip. According to (\ref{dsc}) and (\ref{Phis}), it is straightforward to
see that the phase difference in ${\Delta }$ between two edges across the
strip will change by $2\pi$ (phase slip) for an infinitely long strip.
Hence the boundaries of the sample can always perceive the motion of spinon
vortices inside, which is a direct indication of the phase rigidity. 
The presence of such a new elementary excitation is the most important
feature distinguishing the spontaneous vortex phase from an ordinary normal
state with strong superconducting fluctuations. 

A phase slip of $2\pi $ in a Josephson junction represents an
elementary resistivity process as a {\em voltage} will be generated in that
(transverse) direction\cite{josephson}. 
In an equilibrium state at $T_c<T<T_v $, phase slippage 
is random as those free spinon vortices move in
arbitrary directions (this is an another way to understand the destruction
of the phase coherence in this regime) and the average voltage is zero. In
order to see a {\em net} phase slippage being added up, 
one has to let all the
spinon vortices move in the same direction, say, by maintaining a 
temperature gradient and
use a perpendicular magnetic field to induce an imbalance between vortices and
antivortices. This results in the well known Nernst effect\cite{ri}. The
existence of a nontrivial contribution from spinon vortices to the Nernst
signal, is one of the main features of the spontaneous vortex phase.

There is another way to perceive this effect. In the superconducting
phase, it is well known that the motion of magnetic vortices generates
the Nernst signal. In the present case, as discussed in Ref.\cite{muthu}, a
magnetic vortex is formed by a spinon bound to a magnetic flux
quantized at $hc/2e$ (equal to $\pi $ in the present units$)$ [Fig. 2(a)].
This can be easily seen based on (\ref{dsc}) which shows that each $2\pi $
vortex in the superconducting order parameter must be associated with a
spinon through $\Phi ^s.$ In the bulk, the spinon vortices have to be paired
and do not contribute to $\Phi ^s$ and the net Nernst effect. So
the Nernst signal from $\Phi ^s$ arises only from those spinon vortices
that are nucleated by the magnetic fluxoids and are bound to the latter as
shown in Fig. 2(a). 

At $T_c,$ the phase coherence is destroyed as $\Phi ^s$ becomes disordered
in (\ref{dsc}) due to the unbinding of spinon vortices, 
and $\langle e^{i\Phi ^s(%
{\bf r})}e^{-i\Phi ^s({\bf r}^{\prime })}\rangle $ falls off exponentially
at large distance. Correspondingly, the magnetic vortices collapse and the
Meissner effect disappears; the spinons originally trapped to the
magnetic vortex cores are released from the latter and become free\cite
{muthu}, as illustrated in Fig. 2(b). Since, in the present theory, 
the spinon vortices contribute to the Nernst signal through the order
parameter phase $\Phi ^s$, the collapse of magnetic vortices themselves
does not directly lead to a diminishing Nernst signal above $T_c$. On the
other hand, unbinding vortices and vortices do not give rise to 
{\em additional } Nernst contribution since the net imbalance of vortices and
antivortices 
\begin{figure}[ht!]
\epsfxsize=7.0 cm
\epsfysize=7.5 cm
\centerline{\epsffile{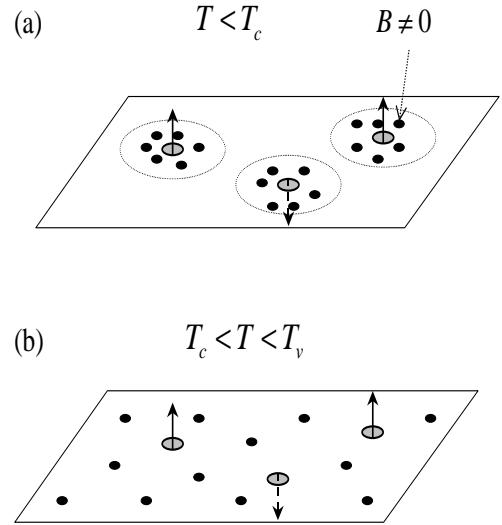}}
\vspace{2mm}
\caption{(a) At $T<T_c,$ a magnetic vortex comprises a spinon vortex
(disks with arrows) and an $hc/2e$ magnetic flux (small full circles  within
the large dashed circles represent finite strength of the
magnetic field); (b) At $T_c<T<T_v,$ magnetic vortices collapse
such that the magnetic field uniformly penetrates through, but
spinon vortices remain.}
\label{fig:2}
\end{figure}
remains the same below and above $T_c$, determined by the
condition 
\begin{equation}
\oint {\bf J}({\bf r})\cdot d{\bf r=0}
\end{equation}
or 
\begin{equation}
\oint d{\bf r}\cdot {\bf A}^e=-\oint d{\bf r}\cdot {\bf A}^s.  \label{flux}
\end{equation}
The only difference is that below $T_c$, the above condition holds for a
closed loop encircling a magnetic fluxoid and far away from the core. (It
also leads to flux quantization at $hc/2e$ as discussed in Ref.\cite
{muthu}.) Above $T_c$, without magnetic vortices, it holds on a spatial
average. Thus the Nernst signal arises from
the
same spinons trapped at magnetic vortex cores below $T_c$ [Fig.
2(a)] and released in the spontaneous vortex phase [Fig. 2(b)], above
$T_c$. Therefore, we expect the
Nernst effect evolve smoothly between the
superconducting and spontaneous vortex phases at $T\rightarrow T_c.$

This smooth evolution of the Nernst effect is probably the most striking
feature observed in the recent Nernst measurement on LSCO compounds by Xu 
{\it et al.}\cite{ong}. These measurements clearly reveal a continuous 
crossover of
the Nernst signal above $T_c$; it remains anomalously enhanced up to 
temperatures, $50-100$ K
above $T_c$. Of course, experimentally one has to separate the vortex
contribution to the Nersnt signal from that of normal charge 
carriers whose contribution is
limited by the so called Sondheimer cancellation\cite{ong}. 
Xu {\em et al.} \cite{ong} argue that the Nersnt signal arises from
vortex like excitations
in the normal state and are not likely to be related to the conventional
superconducting fluctuations. 
Similar phenomena have also been found in
YBCO and BSCO systems as well\cite{ong2,ong1}, 
indicating that the existence of
vortices above $T_c$ is a generic feature of the high-$T_c$ cuprates.

\subsection{Transport coefficients}

\subsubsection{Spinon vortices}

Spinons do not carry charge and thus do not couple to the external
electromagnetic field ${\bf A}^e$ directly. However, they are vortices and
thus generate a Nernst voltage in the direction transverse to their
motion, due to the phase slip effect discussed above. This
can be quantified as follows. In terms of (\ref{jc}), one has 
\begin{equation}
\partial _t{\bf J}=-\frac{\rho _h}{m_h}\left( \partial _t{\bf A}^s+\partial
_t{\bf A}^e\right) .
\end{equation}
The electric field is given by ${\bf E}=-\partial _t{\bf A}^e$ in the
transverse gauge (with $\nabla \phi _h$ being absorbed by ${\bf A}^e,$ which
is always possible below $T_v$).
In the steady state, $\partial _t {\bf J}=0$, and we get
\begin{equation}
{\bf E}=-\sum_l\widehat{{\bf z}}\times {\bf v}^s(l){\alpha _l}\delta ({\bf r}%
-{\bf r}_l)  \label{ey}
\end{equation}
where $l$ labels the spinon vortices with $\alpha _l=\pm \pi $ denoting the
vorticity. Equation (\ref{ey}) means that a finite drift velocity $v_x^s$ of a spinon-vortex
along $\hat{x}$-direction indeed will induce an electric voltage along the $%
\hat{y}$-direction, as the result of the phase slip effect shown in
Fig. 1. Using the condition (\ref{flux}), we get
\begin{equation}
{\bf B}=-\widehat{{\bf z}}\sum_l{\alpha _l}\delta ({\bf r}-{\bf r}_l)
\label{bz}
\end{equation}
where ${\bf B}${\bf \ }denotes the local magnetic field perpendicular to the
2D plane. If all vortices have the same drift velocity, we get
${\bf E}={\bf B}\times {\bf v}^s $
which coincides with the
familiar form for magnetic fluxoids in the flux flow region of a type II
superconductor.

The Nernst coefficient is defined as 
\begin{equation}
\nu _{s-v}=\frac{E_y}{(-\nabla _xT)B_z}.  \label{nuv0}
\end{equation}
Note that in this measurement $J_y$ is set to be zero. $\nu _{s-v}$ can be
determined phenomenologically.
Suppose $s_\phi $ is the transport entropy carried by a spinon vortex and $%
\eta _s$ is its viscosity. Then the drift velocity ${\bf v}^s$ can be
decided by\cite{ong,ri} 
\begin{equation}
s_\phi \nabla T=-\eta _s{\bf v}^s,\label{entropy}
\end{equation}
and consequently, 
\begin{equation}
\nu _{s-v}=\frac{s_\phi }{\eta _s}.  \label{nu2}
\end{equation}
The main result of the bosonic RVB theory is that such an expression is
meaningful {\em both} above and below $T_c$. As discussed earlier,
$\nu _{s-v}$
evolves smoothly between the superconducting and spontaneous vortex
phase at $T\rightarrow T_c$, since the Nernst signal arises from 
the spinon vortices in (\ref{ey}), and the formation or collapse of magnetic
vortices does not change the expression (\ref{nu2}). Of course, below $T_c$, 
$s_\phi $ will include additional contributions associated with the
magnetic vortex core, which remains to be investigated.

Thus, the existence of spinon vortices provides a natural explanation for the 
Nernst experiments
in the spontaneous vortex phase. In this case, the Nernst effect directly
measures the dynamics of spinon vortices as governed by (\ref{hsv}).
The viscosity, $\eta _s$, of spinon vortices is proportional 
to
the scattering rate of the spinons. The latter is enhanced above $T_c$ 
due to
the broadening of the spin resonance level at $E_g$, and
correlated with the fate of the resonance peak observed 
in inelastic neutron scattering
measurements\cite{bourges}. As will be discussed in the next
subsection, $\eta _s$ is also related to the resistivity. 

The transport entropy $s_\phi $, is 
related to the degeneracy of the level $E_s$ as the ``mid-gap'' 
states of the  spinon trapped inside the magnetic vortex core\cite{muthu}.
We expect it to decrease rapidly above $T_c$. This is not due to the collapse 
of the bound core states (of the trapped spinon), 
as in the conventional superconductors. It is due to the 2D Coulomb interaction [(\ref{hv})],
whose effect is important in the high (vortex) density limit.
Note that the degeneracy of the spinon states at the level $E_s$ 
is related to different cyclotron orbits in the LLL (see Sec. IIB). 
In the ``core touching'' 
picture\cite{muthu}, the number of excited spinons at $T_c$, 
is approximately equal 
to the holon number such that each LLL state, 
corresponding to the resonance peak at $E_g,$ is
occupied approximately by one spinon on average. Higher temperature only
means more spinon vortices will be excited to the LLLs such that 
more than one spinon will be found within a cyclotron orbit.
In this dense limit, those 
configurations with many vortices (of the same vorticity) 
lumped together, which 
are allowed by the bosonic statistics, are {\em forbidden} due to costing too 
much potential energy. Thus, the phase space of spinons will be strongly 
limited in the dense limit,
as compared to the dilute case, due to the vortex-vortex interaction, and the entropy 
associated with each spinon vortex should be largely reduced 
above $T_c$ such that the Nernst coefficient 
$\nu _{s-v}$ would decrease in magnitude rapidly. 
A microscopic study based on the effective Hamiltonian (\ref{hsv}) 
is needed in
order to understand this issue and make quantitaive comparison with
experimental results.

Spinon vortices also contribute to the thermal current established by a
temperature gradient, 
\begin{equation}
{\bf J}^Q=\kappa _{s-v}(-\nabla T)  \label{jq0}
\end{equation}
Here the heat current is given by
\begin{equation}
{\bf J}^Q=\sum_{m\sigma }{\bf v}^s(m)E_mn_{m\sigma }^\gamma \approx {\bf v}%
^sE_sn_v  \label{jq}
\end{equation}
with $n_v=\sum_{m\sigma }n_{m\sigma }^\gamma $. In obtaining the last step
in (\ref{jq}), it has been assumed that spinon vortices are mainly excited
to the lowest discrete energy level at $E_s$ and the interaction among
vortices in (\ref{hv}) only gives rise to level broadening which is averaged
out in (\ref{jq}) with ${\bf v}^s$ representing an average drift velocity.

Based on (\ref{entropy})-(\ref{jq}), 
the following connection between the 
thermal conductivity and Nernst effect of the spinon vortices 
can be deduced:
\begin{equation}
\frac{\kappa _{s-v}}{\nu _{s-v}}=\left( \frac {E_g}2 \right)n_v .
\label{nuv1}
\end{equation}
At $T_c$, one has $n_v\sim \delta a^{-2}$ $\cite{muthu}$, and for 
optimally doped YBCO 
with $E_g\sim 41$meV and $\delta\sim 0.15$, we estimate  
$\kappa _{s-v}/\nu _{s-v}\sim 3 $meV$/a^{2}$.

Finally, as a prediction, we introduce
\begin{equation}
\zeta_{s-v}\equiv J^Q_x/{E_yB_z} 
\end{equation}
in the following experiment setting: apply an electric field 
along ${\hat y}$-direction 
and measure the heat current along ${\hat x}$-direction in the 
presence of a perpendicular 
magnetic field $B_z$. In such an experimental situation, 
we expect that spinon vortices contribute predominantly to
$J^Q_x$.
No contribution from phonons will be generated by $E_y$, 
while the Hall effect and the thermal conductivity of the charge carriers, 
holons, are small in the spontaneous vortex phase as will be 
discussed shortly. In this case, 
an electric current induced by $E_y$ along the 
$\hat{y}$-direction exerts a Lorentz force on the 
vortices causing them to move in the $\pm\hat{x}$-directions,
depending on their vorticities 
(see discussion in the next subsection). 
Then, (\ref{ey}) and (\ref{bz}) can be rewritten as 
$E_y=-v_x^s\pi n_v$ and $B_z=-\pi\Delta n_v$, respectively, 
where $\Delta n_v$ denotes the difference
between the numbers of positive and negative vortices. 
The thermal current $J_x^Q=v_x^sE_s\Delta n_v$, 
is related to $\Delta n_v$ instead of $n_v$, 
because vortices with 
opposite vorticities move in opposite directions. Then, we obtain
\begin{equation}
\zeta_{s-v}=\frac{2E_g}{\Phi_0^2n_v},
\end{equation}
where $\Phi _0/2\equiv hc/2e$ 
(full units have been restored).
This quantity is independent of the transport entropy and 
viscosity, and is only related to the 
characteristic spin excitation energy scale and 
the number of spinon vortices.   

\subsubsection{Holons: Charge carriers}

The bosonic holon carries charge $+e$ and directly couples to the
electromagnetic field in (\ref{hh}). In the spontaneous vortex phase, 
holons
are Bose-condensed. But the
``superfluid'' density does not carry zero resistivity in the spontaneous
vortex phase. Rather, one expects to see a {\em finite} resistivity from the
condensed holons, which arises from the motion of spinon vortices.
Suppose an electric current is established along the 
$\hat{x}$-direction. Each spinon vortex carries a $\Phi _0/2$ fictitious
flux with a radius $a_c$ in the holon Hamiltonian (\ref{hh}). Like a
conventional magnetic vortex in a type II superconductor, a transverse
``Lorentz force'' due to the interaction of the current with
the fluxoid deflects the spinon vortex along the $\pm \hat{y}$
directions, depending on the sign of vorticity. In turn, the {\em moving}
spinon vortex produces a voltage along the 
$\hat{x}$ direction, as can be seen from (\ref{ey}). It is then easy to
obtain
\begin{equation}
\rho =\frac{n_v}{\eta _s}\left( \frac{\Phi _0}{2c}\right) ^2 . \label{rho}
\end{equation}

On the other hand, due to the condition (\ref{flux}),
the magnetic field seen by holons will be canceled out in (\ref
{hh}) on the average by ${\bf A}^s$. Thus the Hall resistivity will be
reduced in the spontaneous vortex phase. By the same token, the
magnetoresistance should also be weak, as holons do not see a {\em net}
field on an average. Note that $\rho$ is not sensitive to how 
vortices and
antivortices get polarized by a magnetic field. As seen from  
(\ref{rho}), the resistvity depends only on the {\em total} 
number of excited spinon
vortices, $n_v,$ which as a function of $T$ and $E_s$ should not be very
sensitive to the magnetic field, at least in the weak field limit. 
Finally,
due to the fact that the holon condensate does not carry an entropy, the
thermopower is also expected to be suppressed below $T_v$ and only the
normal-fluid component has a residual contribution.

Transport measurements\cite{thanks} of many cuprate 
superconductors show that 
the suppression of the Hall effect and thermopower\cite{pans,private} 
around
the same temperature scale ($T_v$) above $T_c$ 
where the vortex-induced Nernst effect ends\cite{private}.
These observations are complemented by a weak magnetoresistance 
in the same regime. All of 
this lends a support for the existence of the spontaneous 
vortex phase below $T_v$, 
which can provide a consistent picture as discussed above. 

So far, we have not discussed the role of the nodal (fermionic)
quasiparticle. In the bosonic RVB theory, the quasiparticle arises as a
composite object of a confined holon-spinon pair, with a $d$-wave
dispersion
\cite{muthu0} like in a 
$d$-wave BCS superconductor. But above $T_c$, due to the 
deconfinement of vortices and antivortices, 
such a quasiparticle is expected to be damped,
as its spinon constituent can get away freely. 
Thus, except for residual effects, 
%quasiparticles are no
%longer considered as meaningful elementary excitations in 
%the spontaneous vortex phase. 
their contribution to 
transport should become negligible for $T>T_c$. 
In contrast, the contribution of the quasiparticles becomes dominant
below $T_c$, when their coherence is established. 
\begin{figure}[ht!]
\epsfxsize=7.0cm
\epsfysize=7.5cm
\centerline{\epsffile{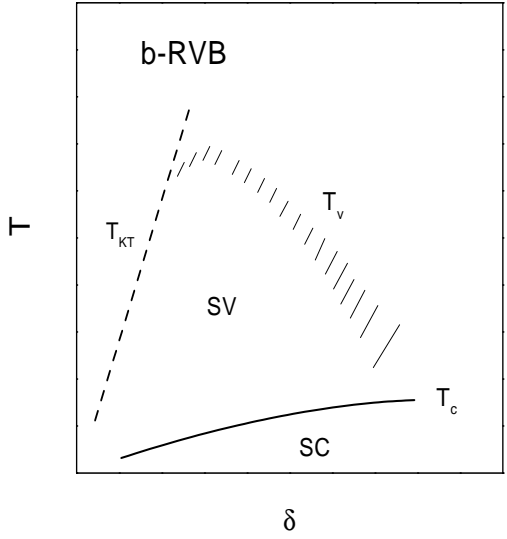}}
%\vspace{2mm}
\caption{Phase diagram. SC -- superconducting state; SV - spontaneous vortex
state. Both SC and SV occur on a bosonic RVB (b-RVB) background ($\Delta
^s\neq 0)$ characterized by short-range antiferromagnetic correlations. }
\label{fig:3}
\end{figure}

\section{Phase diagram and discussions}

The contribution of spinon vortices to the Nernst effect 
vanishes at $T_v$,
when holon condensation or $\Delta _0$ disappears. 
In this section, we discuss the evolution of the spontaneous vortex
phase as a function of doping and temperature and propose a phase
diagram.

Physically, holon condensation at $T_v$
can be interpreted as the original {\em neutral} RVB pairs at high
temperature acquiring charge and becoming Cooper-like pairs. 
Above $T_v$, incoherent 
holons are strongly scattered by the spinon flux-tubes 
according to (\ref{hh}). 
So even though the motion of an RVB spin pair may be accompanied 
a pair of holons as a backflow, it does not 
mean that the RVB pair is ``charged'' by -2e.
The 2e holon backflow becomes coherent only below $T_v$,
when the two holons constituting a pair are condensed. In this regime, 
the RVB spin pair looks like carrying a charge -2e (with a 
{\em single} wavefunction). This is the physical 
picture of how the amplitude of the pairing parameter forms below $T_v$.

The dashed line in Fig. 3 shows the bare 
KT temperature ${T_{KT}}$ for the holon 
condensation, which is obtained without including 
${\bf A}^s$ in (\ref{hh}). First, 
we argue that $T_v\sim T_{KT}$ at very low doping concentrations. 
To understand the 
effect of fluctuating ${\bf A}^s,$
let us first recall the physical interpretation\cite{muthu} of the $T_c$
curve in Fig. 3. The superconducting transition occurs due to the
dissolution of spinon vortex-antivortex pairs in (\ref{dsc})$.$ Such a
transition is not driven by the entropy reason as in a conventional KT
transition. Rather, it is related to the so-called ``core touching''
mechanism: each spinon vortex has a finite core with the length scale
compatible to $a_c$\cite{muthu}, and an upper limit of $%
T_{c}$ is determined by the temperature at which the cores of
excited spinon-vortices start to touch, which leads to $T_c\sim E_g/4$\cite
{ming,muthu}. Considering that the overlapping of vortex cores 
effectively smear out 
the flux fluctuations in (\ref{as1}), 
in spite of unbinding spinon vortices, 
the fluctuations of ${\bf A}^s$ are not necessarily strong 
above $T_c$. This is 
particularly true at small doping when $a_c$ can become very large.
Hence, one expects that $T_v$ approximately coincides with the bare KT
temperature, $T_{KT}=\pi \delta (2a^2m_h)^{-1}$, for small doping.

With the increase of doping, 
the core scale gets reduced. Near optimal doping, 
$\delta =0.15,$ $a_c$ $\sim 1.5a$ which becomes comparable
to the lattice constant. At such a short length scale, core overlapping
no longer results in the smoothness of ${\bf A}^s$. Instead, the
uncorrelated spinons above $T_c$ gives rise to strong flux
fluctuations ($\pm \pi $) through ${\bf A}^s$, effectively
suppressing the holon condensation in (\ref{hh}). Thus 
in this regime, we expect $T_v$ to
be reduced to $T_c$, as shown
in Fig. 3. 
In particular, since $a_c$ also determines the equal time
spin-spin correlation length scale\cite{zy}, at $\delta =0.30$, 
one has $a_c$ $\sim a$, which 
should set an upper doping limit for the bosonic RVB phase, 
as $\Delta ^s$ is associated
with nearest-neighbor antiferromagnetic correlations.
Both $T_v$ and $T_c$ are expected to vanish at 
the point where the 
bosonic RVB order parameter $\Delta ^s$ disappears on the 
$T=0$ axis.  

The phase diagram of LSCO has been mapped out carefully, based on
high resolution Nernst experiments\cite{ong1}. 
For doping
concentrations between 0.03 and 0.07, 
the experimental $T_v$ (denoted by $T_\nu $
in Ref.\cite{ong1}) increases very steeply from 
0 to $90$ K with a slope
$>2,400$ K, which puts $(m_ha^2)^{-1}>0.13$ eV in the present theory. 
The
experimental $T_v$ then peaks around $\sim 128$ $K$ at $\delta =0.11$ and
decreases monotonically at larger doping, a trend similar to the plot shown
in Fig. 3.

It is noted that the sharpness of
the resonance-like peak at $E_g$ is also caused by the holon condensation
in the present theory\cite{zy}. Thus, $T_v$ is intrinsically related
to the spin ``pseudogap'' temperature $T^{*},$ below which the resonance
peak starts to sharpen up in neutron-scattering measurements\cite{bourges}
and the spin-lattice relaxation rate $1/T_1$ begins to deviate from the
high-T non-Korringa behavior in NMR measurements. In the transport channel,
the ``pseudogap'' seen in resistivity should be also related to the holon
condensation. Similarly, the holon condensation plays an essential role in
the single-particle channel\cite{muthu0} which may explain the ``pseudogap''
seen in photoemission spectroscopy measurements. But it is important to
point out since no {\em true} phase transition takes place and these
temperature scales do not necessarily coincide with each other. They 
represent crossover temperatures in different channels in response to
the holon condensation or the forming of the pairing amplitude $\Delta _0$.
For instance, the transport $T^{*}$ is generally in a higher curve,
indicating the {\em onset} of holon coherence before the holon condensation.
If this picture holds true in the cuprates, 
then one also expects $T_v$ to behave
like the shaded curve in Fig. 3, which quickly decreases towards $T_c$ such
that the spontaneous vortex phase shrinks as the doping concentration $
\delta $ approaches the optimal and over doped regimes.

In summary, we have investigated the normal state below $T_v,$ the onset
temperature for the {\em amplitude }of Cooper pairs, based on 
the bosonic RVB theory. Such a phase is characterized by spinons
that behave as {\em free vortices}. $T_v$ coincides with the {\em holon}
condensation, while $T_c$ is {\em lower} at which the phase coherence is
realized due to the binding of spinon-vortices and -antivortices. In the
spontaneous vortex phase, the transport properties are quite unique. We
showed that free spinon vortices contribute to the Nernst effect, 
which evolves smoothly into the superconducting phase, 
consistent with the Nernst
experiments. We also argued that the holon condensation is
responsible for the pseudogap phenomena and that $T_v$ is correlated 
with various
``pseudogap'' temperature scales observed experimentally. The bosonic
RVB theory of the $t-J$ model offers a framework to unify many of the
seemingly disparate phenomena observed in the high $T_c$
superconductors, and we hope to address these in the future.

\acknowledgments 
We thank P. W. Anderson and N. P. Ong for very useful discussions. Z.Y.W.
thanks the hospitality of Physics Department at Princeton University where
this work was finished, and acknowledges partial support from NSFC Grant
90103021. V. N. M. acknowledges partial support from NSF Grant DMR 98-09483.
We thank N. P. Ong and Yayu Wang for sharing their unpublished
results with us.
\appendix{}

\section{Spin rotational invariance}

According to the holon effective Hamiltonian (\ref{hh}), the bosonic spinons
are perceived by holons as $\pi $-{\em flux-tubes}, represented by ${\bf A}%
^s $ defined in (\ref{as1}). Since in the definition (\ref{as1}), the {\em %
sign} of the flux depends on the {\em spin index}, one may naturally raise
the question as whether the spin rotation may be broken in such a system. In
the following we give an explicit proof that the spin rotational symmetry is
retained.

For convenience, we will use the original lattice version of $H_h$\cite{zy}: 
\begin{equation}
H_h=-t_h\sum_{\langle ij\rangle }\left( e^{i[A_{ij}^e+A_{ij}^s-\phi
_{ij}^0]}\right) h_i^{\dagger }h_j+H.c.  \label{hh0}
\end{equation}
where $A_{ij}^e$ represents the external electromagnetic field, and the
lattice version $A_{ij}^s$ is defined by 
\begin{equation}
A_{ij}^s=\frac 12\sum_{l\neq i,j}{\mbox {Im}}\ln \left[ \frac{z_i-z_l}{%
z_j-z_l}\right] \left( \sum_\sigma \sigma n_{l\sigma }^b\right)  \label{eas}
\end{equation}
which describes fictitious fluxoids bound to spinons satisfying 
\begin{equation}
\sum_cA_{ij}^s=\pm \pi \sum_{l\in c}\left[ n_{l\uparrow }^b-n_{l\downarrow
}^b\right]
\end{equation}
for a closed loop $c$. $\phi _{ij}^0$ corresponds to a uniform $\pi $ flux
per plaquette.

In the phase string formulation\cite{zy}, the spin operators are defined as follows: 
\begin{equation}
S_i^z=\sum_\sigma \sigma b_{i\sigma }^{\dagger }b_{i\sigma },
\end{equation}
\begin{equation}
S_i^{+}=b_{i\uparrow }^{\dagger }b_{i\downarrow }(-1)^ie^{i\Phi _i^h},  \label{s+}
\end{equation}
and $S_i^{-}=(S_i^{+})^{\dagger }.$ Here

\begin{equation}
\Phi _i^h=\sum_{l\neq i}%
%TCIMACRO{\func{Im} }
%BeginExpansion
\mathop{\rm Im}%
%EndExpansion
\ln \left[ z_i-z_l\right] n_l^h.  \label{phih}
\end{equation}

It is obvious that

\begin{equation}
\left[ H_{h,}\text{ }S_l^z\right] =0.
\end{equation}
On the other hand, one finds that the phase $\Phi _i^h$ defined in (\ref
{phih}) plays a crucial role in compensating the extra phase generated from $%
A_{ij}^s$ by a spin flip, which results in 
\begin{equation}
\left[ H_{h,}\text{ }S_l^{\pm }\right] =0.
\end{equation}
Also note that if the summation in $H_h$ involves the links $ij$ which
coincide with $l,$ such terms have no contribution due to the
no-double-occupancy constraint.

So generally one has 
\begin{equation}
\left[ H_{h,}\text{ }{\bf S}_i\right] =0
\end{equation}
which means that the holon effective Hamiltonian $H_h$ in (\ref{hh0}) or (%
\ref{hh}) not only always respects the {\em global} spin rotational
symmetry, but also respects a local spin symmetry. This is consistent with
the notion of spin-charge separation in general, and that the spin index is
independent of the vorticity of a spinon-vortex composite, discussed in Sec.
IIA, in particular. In other words, $\pi $-flux-tubes bound to spinons, seen
by holons, and the current-vortices bound to spinons only reflect the fact
of electron{\em \ fractionalization}, without involving any spin
symmetry-breaking. To understand this, one has to realize that there is a
peculiar formulation of $S_i^{\pm }$ [(\ref{s+})] in the phase string
formalism, as shown above.


\begin{references}
\bibitem{V-K}  V. J. Emery and S. A. Kivelson, Nature {\bf 374}, 434(1995).

\bibitem{anderson}  P. W. Anderson, Science {\bf 235}, 1196 (1987).

\bibitem{baskaran}  G. Baskaran, Z. Zou, and Anderson, Solid State Commun. 
{\bf 63}, 973 (1987).

\bibitem{muthu}  V. N. Muthukumar and Z. Y. Weng, cond-mat/0112339.

\bibitem{zy}  Z. Y. Weng, D. N. Sheng, and C. S. Ting, 
Phys. Rev. Lett. {\bf 80}, 5401 (1998); 
Z. Y. Weng, D. N. Sheng, Y. C. Chen, and 
C. S. Ting, Phys. Rev. B {\bf 55}, 3894 (1997).

\bibitem{remark2}  For simplicity, the d-wave prefactor is dropped here.

\bibitem{ong}  Z. A. Xu, N. P. Ong, Y. Wang, T. Kakeshita, and 
S. Uchida, Nature, {\bf 406}, 486 (2000).

\bibitem{ong2}  Y. Wang, Z. A. Xu, N. P. Ong, and S. Uchida, unpublished.

\bibitem{ong1}  Y. Wang, Z. A. Xu, T. Kakeshita, S. Uchida, S. Ono, 
Y. Ando, and N. P. Ong, cond-mat/0108242.

\bibitem{bourges}  {\em see}, P. Bourges, in {\em The Gap Symmetry and
Fluctuations in High Temperature Superconductors}, 
Ed. J. Bok, {\em et al}. (Plenum Press, 1998).

\bibitem{ming}  Ming Shaw, Z.Y. Weng, and C.S. Ting, cond-mat/0110527.

\bibitem{josephson}  B. D. Josephson, Phys. Lett. {\bf 16}, 242 (1965);
P. W. Anderson, Rev. Mod. Phy. {\bf 38}, 298 (1966).

\bibitem{ri}  H.-C. Ri, R. Gross, F. Gollnik, A. Beck, 
R.P. Huebener, P. Wagner, and H. Adrian, 
Phys. Rev. B {\bf 50}, 3312 (1994).

\bibitem{thanks} N. P. Ong, private communication.

\bibitem{pans}  Y. Wang and N. P. Ong, PNAS {\bf 98}, 11091 (2001); 
Z. A. Xu, Y. Zhang, and N.P. Ong, cond-mat/9903123.

\bibitem{private} N. P. Ong, unpublished.

\bibitem{muthu0} Z.Y. Weng, D.N. Sheng, and C.S. Ting, 
Phys. Rev. B {\bf 61}, 12328 (2000); 
V. N. Muthukumar, Z.Y. Weng, and D.N. Sheng, cond-mat/0106225.
\end{references}
\end{document}